\documentclass[showpacs,epsf,twocolumn]{revtex4-1}
\usepackage{float}
\usepackage{color,graphicx}
\graphicspath{ {images/} }
\usepackage{hyperref}

\begin{document}

\title{Effect of hydrostatic pressure on ferromagnetism in two-dimensional CrI$_3$}

\author{Suchanda Mondal$^1$, Murugesan Kannan$^2$, Moumita Das$^1$, Linganan Govindaraj$^2$, Ratnadwip Singha$^1$, Biswarup Satpati$^1$, Sonachalam Arumugam$^2$ and Prabhat Mandal$^1$}

\affiliation{$^1$Saha Institute of Nuclear Physics, HBNI, 1/AF Bidhannagar, Calcutta 700 064, India}
\affiliation{$^2$Centre for High Pressure Research, School of Physics, Bharathidasan University, Tiruchirappalli 620 024, India}

\date{\today}

\begin{abstract}
We have investigated the magnetic properties of highly anisotropic layered ferromagnetic semiconductor CrI$_3$ under hydrostatic pressure with magnetic field along the easy-axis of magnetization. At ambient pressure, CrI$_3$ undergoes a second-order paramagnetic to ferromagnetic phase transition at $T_C$=60.4 K. $T_C$ is found to increase sublinearly from 60.4 to 64.9 K as pressure increases from 0 to 1.0 GPa. With the increase in pressure, the transition becomes sharper while magnetization at low-field decreases monotonically due to the decrease in magnetocrystalline anisotropy. The weak low-field anomaly at around 48 K, resulting from the two-step magnetic ordering, also shifts toward higher temperature with increasing pressure. The observed increase in $T_C$ and the decrease in magnetization could originate from change in coupling between the layers and Cr-I-Cr bond angle with pressure.
\end{abstract}
\maketitle

Two-dimensional ferromagnetic (FM) semiconductors exhibit a wide range of novel electronic properties with immense potential for application in different magnetoelectronic technology \cite{bur,park,hung,hell}. In spintronic devices, instead of or in addition to charge degrees of freedom, electron spin is used for information storage and processing \cite{baib,bina,jull,datta,zhong}. This can be realized, when magnetism is incorporated into the active materials. Low-dimensional magnetic systems with intrinsic ferromagnetism have also generated considerable interest for fundamental research due to the occurrence of several quantum phenomena \cite{onsa,mer,bere,kost,col,siva}. According to the Mermin-Wagner theorem, two-dimensional (2D) isotropic Heisenberg model does not show long-range magnetic ordering at any finite temperature \cite{mer}. On the other hand, a 2D Ising system does order at finite temperature, as predicted by Onsager in his remarkable paper \cite{onsa}. Recently, two different types of layered FM semiconductors with weak inter-layer van der Waals interaction, namely, chromium trihalides, Cr$X_3$ ($X$$=$Cl, Br, I), and chromium-based ternary compounds, Cr$_2$$Z$$_2$Te$_6$ ($Z$$=$Si, Ge), have received considerable attention as promising candidates for possible application in spintronic technology owing to their easy exfoliation property and the existence of long-range ferromagnetism across the bilayer or within a monolayer \cite{Mcg,Gong,Huang,liu,mag,wang,web,lado}. Among different members of Cr$_2Z_2$Te$_6$  and Cr$X_3$ series, bulk Cr$_2$Ge$_2$Te$_6$ and CrI$_3$ compounds exhibit highest FM transition temperature ($T_C$), 68 and 61 K, respectively. Though, their $T_C$s are comparable, in contrast to Cr$_2$Ge$_2$Te$_6$, the magnetic interaction is much more anisotropic in CrI$_3$ due to the very weak coupling between the layers. The ferromagnetism is retained in the single layer, with transition temperature as high as 45 K. It is, in fact, very interesting that CrI$_3$ is antiferromagnetic (AFM) in bilayer \cite{Huang}. Also, an extremely high tunneling magnetoresistance (10$^5$ $\%$) could be observed in an exfoliated thin film of CrI$_3$ \cite{wang}. \\

Unlike chemical substitution, hydrostatic pressure ($P$) is a continuously tunable thermodynamic variable which can be used to tune the phase transition as well as charge conduction mechanism without introducing any disorder in the system. Primarily, pressure alters the bond lengths and angles of the lattice which, in turn, affect the intersite electron hopping process. As a result, both transport and magnetic properties may change significantly with application of pressure. For example, several systems undergo pressure-induced insulator to metal transition \cite{saka,past,foro}. Even, exotic phenomena like superconductivity at close to room temperature can be achieved by applying high pressure \cite{droz,soma}. In perovskite manganites, both electronic and magnetic subsystems are found to be highly susceptible to  external pressure as well as chemical pressure arising  out of variation in ionic size \cite{sar,sar1,dem}. However, the role of pressure on $T_C$ in an insulating ferromagnet is a very complicated and debatable issue due to the competition between direct exchange and superexchange. It is believed that the superexchange interaction strength increases with compression and as a result, the pressure coefficient of $T_C$ increases linearly with the isothermal compressibility \cite{kano}. Though, $T_C$ has been found to increase with increase in $P$ in several insulating ferromagnets, there are examples where $T_C$ decreases with applied pressure \cite{kano}. Very recently, magnetic properties of Cr$_2$Ge$_2$Te$_6$ have been investigated under pressure  and $T_C$ was found to decrease monotonically with pressure up to 1 GPa \cite{sun}. Also, a small but negative pressure coefficient of $T_C$ has been reported for CrBr$_3$ \cite{yoshi}. On the other hand, $T_C$ increases above a critical value of applied pressure in layered VI$_3$ \cite{son}. So, it is important to investigate systematically the role of pressure on ferromagnetism in several insulating layered ferromagnets. In the present work, we have studied the effect of hydrostatic pressure on magnetic properties of well characterized single crystalline sample of CrI$_3$. In contrast to Cr$_2$Ge$_2$Te$_6$ and CrBr$_3$, $T_C$ in CrI$_3$ is observed to increase monotonically with pressure. Also, with increasing pressure, the paramagnetic-ferromagnetic (PM-FM) transition becomes sharper and the value of magnetization with field parallel to $c$-axis decreases. \\

Single crystals of CrI$_3$ were grown by standard chemical vapor transport technique using  high purity elements. The analyses of the phase purity and structural details were carried out using both high-resolution x-ray diffraction and  transmission electron microscopy. The magnetic measurements were performed in a 7 T SQUID-VSM. The bulk single crystal of CrI$_3$ undergoes a long-range ferromagnetic ordering below $T_C$$\sim$60.4 K. The details of sample preparation, characterization and magnetic measurements are provided in the Supplementary Materials section \cite{supp}.\\

\begin{figure}
\includegraphics[width=0.5\textwidth]{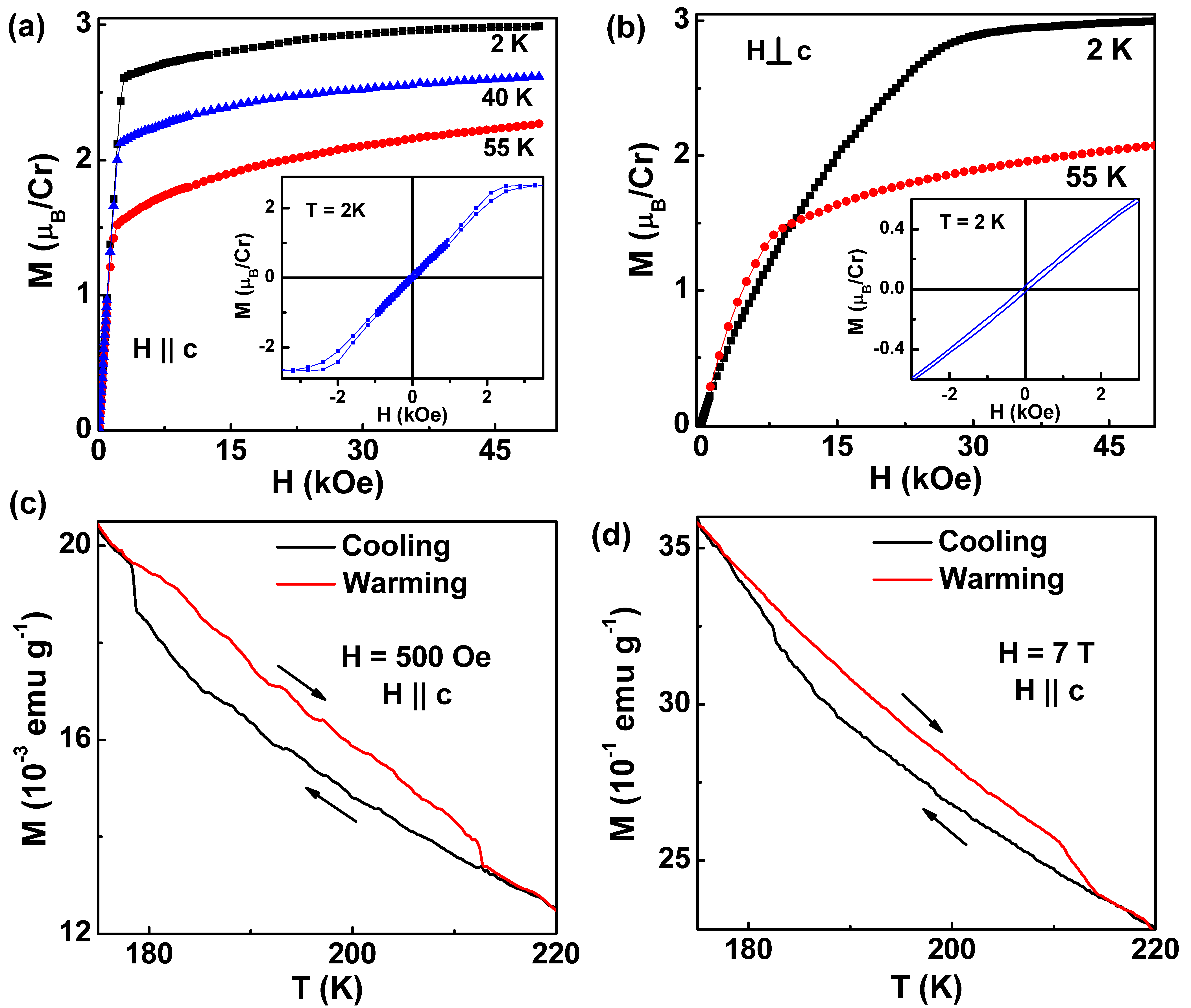}
\caption{(Color online) (a) and (b) Field dependence of magnetization for CrI$_3$ crystal measured at different temperatures with applied field  parallel to $c$ axis and perpendicular to $c$ axis, respectively. Insets show hysteresis loop in the low-field regime. The observed behavior is qualitatively similar to what has been reported earlier \cite{Mcg,liu}. (c) and (d) Field-cooled magnetization for heating and cooling cycles with magnetic field  500 Oe and  7 T, respectively applied along $c$ axis. A clear anomaly along with thermal hysteresis can be seen below $\sim$212 K due to the first-order structural transition.}
\end{figure}

Figures 1(a) and (b) show isothermal magnetization curves at few representative temperatures with field parallel to $c$ axis ($H$$\|$$c$)  and perpendicular to $c$ axis ($H$$\bot$$c$), respectively. Well below the $T_C$, magnetization increases sharply and starts to saturate above 3 kOe for $H$$\|$$c$, whereas,  $M$ approaches saturation at a much higher field for $H$$\bot$$c$, above 30 kOe. This indicates that CrI$_3$ is a highly anisotropic system and $c$ axis is the easy axis of magnetization. Both the magnitude and nature of field dependence of $M$ are very similar to the observations made earlier \cite{Mcg,liu}. Figure 1(a) shows that $M$ decreases monotonically with increase in temperature as in the case of a ferromagnet but the critical field ($H_{sat}$) above which $M$ exhibits saturation-like behavior does not change significantly with temperature. However, $M$ exhibits an anomalous behavior for $H$$\bot$$c$. With increasing temperature, though, $M$ decreases at high fields, it shows a small increase in the low-field region. Figure 1(b) shows that $M$ is slightly higher for 55 K as compared to that for 2 K when the applied field is below 1 T. Above 1 T, $M$ increases slowly. Thus, for $H$$\bot$$c$, $H_{sat}$ decreases rapidly with increasing temperature. Similar behavior has  been observed in layered itinerant ferromagnet Fe$_3$GeTe$_2$ \cite{chen}. This unusual behavior of $M$ is due to the strong suppression of magnetocrystalline anisotropy with increasing temperature \cite{rich}. The value of saturation moment at 2 K and 5 T is about 2.98 $\mu_B$/Cr$^{3+}$ ion, which is very close to the expected moment (3 $\mu_B$) in high-spin configuration state of the Cr$^{3+}$ ion. Insets of Figs. 1(a) and (b) show the hysteresis loop in the low-field region for $H$$\|$$c$ and $H$$\bot$$c$, respectively. In both the cases, hysteresis is very weak. The coercive field is about 72 Oe for $H$$\|$$c$  and 85 Oe for $H$$\bot$$c$, which are close to earlier report \cite{Mcg}. The weak coercivity suggests that CrI$_3$ is a soft ferromagnet similar to CrGeTe$_3$ and CrSiTe$_3$ \cite{sun}.\\

CrI$_3$ exhibits a first-order structural phase transition from monoclinic to rhombohedral below $T_S$ \cite{Mcg}. In order to investigate how this transition affects magnetic properties, magnetization has been measured by cooling the sample from room temperature down to 2 K in presence of field and subsequent warming, as shown in Figs. 1(c) and (d). $M$ displays an anomaly around  $T_S$$\sim$212 K, and the cooling and heating cycles data do not overlap with each other but display a significant thermal hysteresis over a wide  range 212-180 K, which is consistent with the reported temperature dependent x-ray diffraction \cite{Mcg}. Though the value of $M$ changes systematically with $H$, the nature of $M$($T$) curve in the range 212-180 K is found to remain insensitive to field up to 7 T.  We have also performed magnetization measurements with $H$$\bot$$c$ and observed similar behavior.\\

\begin{figure}
\includegraphics[width=0.5\textwidth]{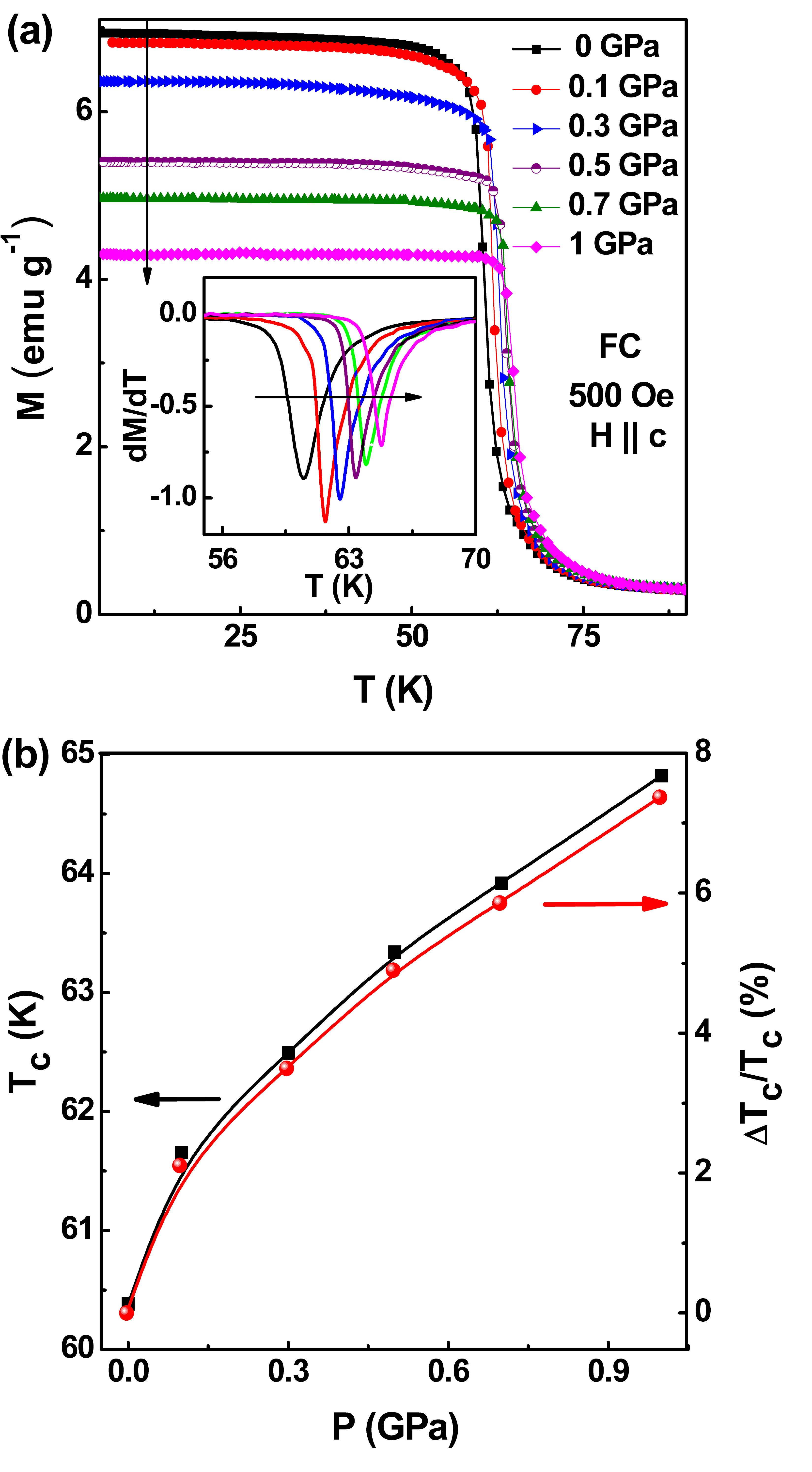}
\caption{(Color online) (a) Temperature dependence of field-cooled magnetization for CrI$_3$ crystal at different applied pressures with magnetic field parallel to the $c$ axis. Inset shows $dM/dT$ as a function of $T$ for different pressures. Arrows indicate the direction of increase in pressure. (b) $T_C$, determined from the position of the minimum in $dM/dT$ curves, along with the relative change in $T_C$ ($\Delta$$T_C/T_C$) are plotted as a function of pressure.}
\end{figure}

To capture the salient features of ferromagnetism under pressure, field-cooled magnetization with $H$$\|$$c$ has been plotted in Fig. 2(a) as a function of temperature for different applied pressures. With increase in $P$, $M$ decreases monotonically and the transition region shifts slowly toward higher temperature. The increase in $T_C$ with increase in $P$ suggests that pressure stabilises the ferromagnetic state. However, the sharp decrease in $M$ with increasing pressure is quite unusual and appears to be inconsistent with this simple picture. Usually, in a simple ferromagnet, magnetic moment is expected to increase when $T_C$ increases with application of pressure. In Cr$_2$Ge$_2$Te$_6$, $M$ shows a non-monotonic dependence on $P$ \cite{sun}. Initially, $M$ increases with pressure and then decreases slowly. $M$ remains slightly smaller at ambient pressure. In order to determine the dependence of $T_C$ on pressure in CrI$_3$, $T_C$ has been estimated from the $dM/dT$ versus $T$ curves.  For each pressure, d$M$($T$)/d$T$ curve shown in the inset of Fig. 2(a), exhibits a very sharp minimum and it shifts progressively toward higher temperature with increasing pressure. Indeed, we observe that the full-width at half-minimum of $dM/dT$ versus $T$ curve is smaller for $P$$>$0.  In particular, the transition is very sharp for 0.1 GPa. Thus, the PM-FM transition in CrI$_3$ remains  sharp with the application of pressure up to 1 GPa. In Cr$_2$Ge$_2$Te$_6$, the transition becomes broader under pressure \cite{sun}. The pressure variation of $T_C$ is demonstrated in Fig. 2(b). Figure shows that the dependence of $T_C$ on pressure is not linear. Initially, $T_C$ increases very rapidly with pressure at the rate of $\sim$12 K/GPa up to about 0.10 GPa and then continues to increase further, albeit, at a slower rate. For $P$$\geq$ 0.5 GPa, $T_C$ increases approximately linearly with a pressure coefficient, $dT_C/dP$$\sim$3 K/GPa. We have also calculated the relative change in $T_C$, defined as $\Delta$$T_C$/$T_C$=$[T_C(P)-T_C(0)]/T_C(0)$, which has been plotted along with $T_C$. This plot shows that $T_C$ increases by about 7.5 $\%$ for an applied pressure of 1 GPa.\\

\begin{figure}
\includegraphics[width=0.5\textwidth]{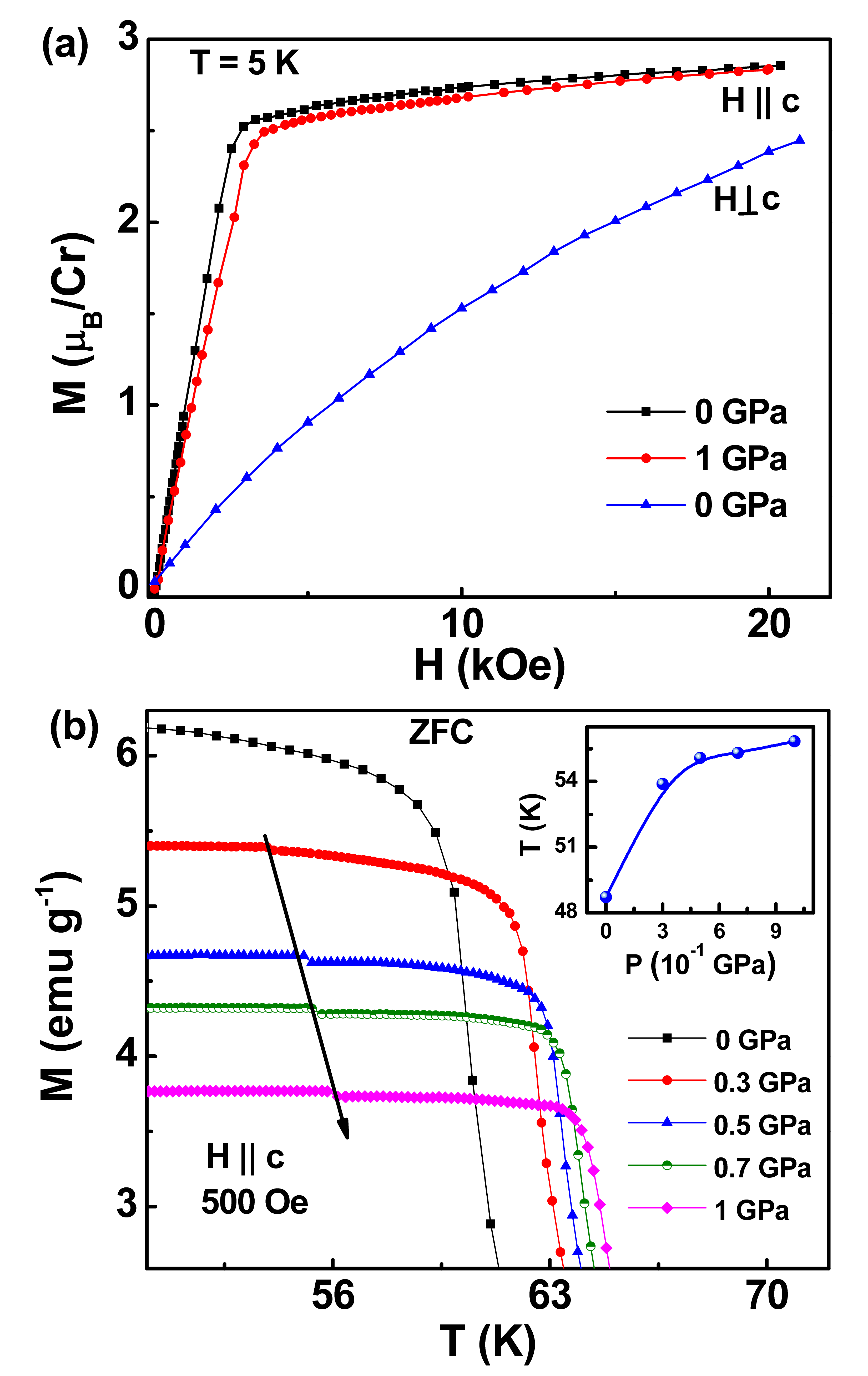}
\caption{(Color online) (a) Field dependence of magnetization curves for CrI$_3$ crystal measured at ambient and 1 GPa pressure with field  parallel and perpendicular to $c$ axis at a representative temperature 5 K. (b) Temperature dependence of zero-field-cooled (ZFC) magnetization for CrI$_3$ at different applied pressures with magnetic field parallel to the $c$ axis. Inset shows the pressure dependence of two-step transition temperature $T^\star$, determined from the position of the weak anomaly in $M$($T$) curves as shown by the arrow in the main panel.}
\end{figure}

In order to understand and explain the systematic decrease of magnetization with applied pressure in Fig. 2(a), we have measured the magnetic field dependence of $M$ below $T_C$ with $H$$\|$$c$ configuration. In Fig. 3(a), magnetic moment under 1 GPa pressure has been plotted as a function of $H$ at temperature 5 K. In the same figure, $M$($H$) curves at ambient pressure for $H$$\|$$c$ and $H$$\bot$$c$ configurations are also shown for comparison.  The magnetization under pressure also increases rapidly with increasing field and starts to saturate above 3 kOe as in the case of ambient pressure with $H$$\|$$c$. Above 1 T, the value of $M$ is  close to that for ambient pressure with $H$$\|$$c$ and at around 2 T both the curves merge with each other. This suggests that the Cr$^{3+}$ remains in the high-spin state under pressure. However, closer view reveals a significant difference between $M$($H$) curves at ambient pressure and 1 GPa. In the low-field regime, the slope of $M$($H$) curve, i.e., the magnetic susceptibility ($dM/dH$), is smaller for 1 GPa than that for the ambient pressure.  The smaller value of $dM/dH$ implies that the value of field required to achieve a state of saturation in magnetization increases with applied pressure for $H$$\|$$c$. This may be due to the decrease in magnetic anisotropy under  pressure in CrI$_3$. The field dependence of $M$ under pressure with $H$$\bot$$c$ could shed some light on this issue. This is, however, not possible with our current setup. If the anisotropy reduces under pressure, the $M$($H$) curves with $H$$\|$$c$ and $H$$\bot$$c$ should approach each other with increasing pressure. At low applied field, $M$($T$) curve exhibits a weak anomaly slightly below $T_C$, which is more prominent in zero-field-cooled cycle with $H$$\bot$$c$. This anomaly in $M$ has been characterized as a two-step magnetic ordering but its origin is yet to decipher \cite{liu1}. We have also recorded the temperature dependence of $M$ in zero-field-cooled cycle at different pressures with $H$$\|$$c$ to track this two-step transition at $T$$=$$T^\star$, as shown in Fig. 3(b). The pressure dependence of $T^\star$ is shown in the inset of Fig. 3(b). Similar to $T_C$, $T^\star$  also increases sharply in the low-pressure region and then increases very slowly.\\

Mainly two competing interactions are responsible for determining the magnetic ground state of CrI$_3$. The direct exchange between Cr-Cr is AFM in nature. This direct exchange arises from the electron hopping between  nearest-neighbor Cr sites and is maximum when the Cr-Cr bond angle, formed by the 3$d$ orbitals, is 180$^\circ$. On the other hand, depending on the symmetry relations and electron occupancy of the overlapping atomic orbitals, superexchange can be FM as well as AFM in nature and is mediated through a non-magnetic ligand ion. Unlike the direct exchange, the superexchange interaction originates due to the virtual hopping of electrons between the two nearest-neighbor Cr ions via iodine ion. This virtual process reduces the total energy of the system. According to the Goodenough-Kanamori-Anderson rules, when the magnetic-ion-ligand-magnetic-ion angle is 90$^\circ$, the superexchange interaction is FM. It is AFM, when the angle is 180$^\circ$ \cite{and,good,kan}.  \\

In a weakly coupled layered van der Waals system, the effect of pressure on ferromagnetism is very sensitive to the bond angles  and the inter-layer coupling. If the magnetic-ion-ligand-magnetic-ion bond angle is 90$^\circ$, $T_C$ is expected to decrease with application of  pressure because pressure will either increase or decrease the bond angle from 90$^\circ$. As a result, the superexchange interaction, which favors the FM ordering, weakens. On the other hand, $T_C$ may increase with pressure, if the bond angle is lower or higher than 90$^\circ$ at ambient pressure. In such a situation, the bond angle may approach toward 90$^\circ$ with the application of pressure. It has already been mentioned that the effect of pressure on ferromagnetism, investigated on bulk single crystal of Cr$_2$Ge$_2$Te$_6$ through magnetization measurements, is very different from what is observed in CrI$_3$ \cite{sun}. In Cr$_2$Ge$_2$Te$_6$, $T_C$ decreases monotonically with pressure up to 1 GPa.  The first-principles calculations show that the Cr-Cr bond length decreases while the Cr-Te-Cr bond angle gradually diverges from 90$^\circ$ with increasing pressure. Both these effects favor the direct exchange and weaken the superexchange interaction, i.e., AFM interaction in Cr$_2$Ge$_2$Te$_6$ enhances under pressure. Theoretical calculation also shows that the $c/a$ ratio decreases with pressure.  The reduction of $c$ is relatively more significant than that of $a$ due to the weak interlayer interaction. As the interlayer coupling in CrI$_3$ is much weaker as compared to that in Cr$_2$Ge$_2$Te$_6$, the effect of pressure in increasing the coupling between two adjacent layers is much more significant in the former. In bulk CrI$_3$, the interlayer coupling ($J$) is FM in nature. Thus, one expects that $T_C$ will increase with the increase in pressure due to the enhancement of interlayer FM coupling. Our $M$($H$) data under pressure also offer evidence of enhancement in interlayer coupling under pressure. In VI$_3$, another ferromagnet with weak van der Waals coupling, $T_C$ increases and the transition becomes sharp above a threshold value of $P$, which has been attributed to a crossover from two- to three-dimensionality due to the increase in interlayer coupling \cite{son}. In order to understand why CrI$_3$ exhibits highest $T_C$ among the Cr$X_3$ family and the effect of pressure on FM transition in CrI$_3$ is different from that in CrBr$_3$, we first briefly discuss the evolution of magnetism in Cr$X_3$ with the size of halogen ion $X$. As the Cr-Cr distance increases with increasing halogen size from Cl to Br to I, the direct exchange weakens.  Also, by moving from Cl to Br to I, the covalent nature of Cr-$X$ bond enhances which further strengthens the superexchange interaction as well as the spin-orbit coupling and hence enhances the FM ordering temperature \cite{lado}. Recent experiments suggest that the covalent nature of Cr-I bond  plays an important role in CrI$_3$ to engineer highest $T_C$ \cite{fris}.\\

In order to take into account the anisotropic nature of FM superexchange interaction via Cr-I-Cr, Lado and Fernandez-Rossier used anisotropic $XXZ$-type Hamiltonian with an additional term in the Heisenberg model, anisotropic symmetric exchange ($\lambda$), and estimated $T_C$ from the spin wave theory \cite{lado}. They observe that $T_C$ increases with the increase in $\lambda$. In fact, the nature of the variation of $T_C$ with $\lambda$ is very similar to what we observe here from the pressure dependence. Another factor that may play an important role to increase the $T_C$ in CrI$_3$ system is the Cr-I-Cr bond angle. In CrI$_3$, this bond angle is about 95$^\circ$ \cite{lado,web}. With increase in pressure, the Cr-I-Cr bond angle may approach toward 90$^\circ$. We believe that the different role of pressure on the FM transition of CrI$_3$ and CrBr$_3$ is partly due to their structural anisotropy and partly due to the strong covalent nature of Cr-I-Cr bonding. Reduction in anisotropy with pressure also signifies crossover from an Ising to a Heisenberg system. This leads to suppression in both $T_C$ and susceptibility. On the other hand, as discussed earlier, the change in bond angle and increase in interlayer coupling $J$ could give rise to enhancement in $T_C$ under pressure. It appears, therefore, that the close competition between these two effects determines the nature of pressure dependence of $T_C$ and magnetization in different 2D magnetic systems. For example, while in CrI$_3$, we have observed drop in low-field magnetization and nonlinear rise in $T_C$, the opposite behavior could be observed in CrBr$_3$ and other similar compounds. High pressure structural analysis may reveal important information for understanding the role of pressure on FM interaction in Cr$X_3$ series.\\

In conclusion, we have studied the magnetic properties of quasi-two-dimensional FM semiconductor CrI$_3$ under hydrostatic pressure. Unlike the two-dimensional FM semiconductors  CrBr$_3$ and Cr$_2$Ge$_2$Te$_6$, in the present system, $T_C$ is found to increase monotonically from 60.4 to 64.9 K as pressure increases from 0 to 1.0 GPa. With application of pressure, the PM-FM transition becomes sharper whereas the low-field magnetization decreases for magnetic field parallel to easy axis of magnetization. $M$($H$) data suggest that pressure makes the Cr$^{3+}$ spins harder to orient along $c$ axis, as a result, low-field magnetization decreases with increase in pressure.\\

Acknowledgement: The authors acknowledge the technical support from A. Paul during sample preparation and the measurements. SA thanks DST (SERB, PURSE, FIST), UGC-DAE-CSR Indore, CEFIPRA, New Delhi. M.K thanks CSIR for the fellowship.

\end{document}